# Exploring Fresnel diffraction at a straight edge with a neural network

C Finot, [1] and S Boscolo [2]

[1]Laboratoire Interdisciplinaire Carnot de Bourgogne, UMR 6303 CNRS - Université de Bourgogne-Franche-Comté, 9 avenue Alain Savary, BP 47870, 21078 Dijon Cedex, France
[2]Aston Institute of Photonic Technologies, School of Engineering and Applied Science, Aston University, Birmingham B4 7ET, United Kingdom

E-mail : christophe.finot@u-bourgogne.fr

**Abstract.** We describe a research project carried out with a group of undergraduate physics students and aimed at exploring the use of a neural network to study a classical problem in wave optics whose analytical solution is well known: the diffraction of light by the straight edge of an opaque semi-infinite screen. Through this exposure to machine learning, the students were able to appreciate the basic steps involved in a machine-learning process.

## 1. Introduction

Wave diffraction is one of the most exciting topics for an undergraduate physics student, but it is also demanding as it relies on mathematical prerequisites. Therefore, a seemingly basic problem such as working out the diffraction pattern that is induced by a semi-infinite obstacle placed in the path of a monochromatic plane wave is not intuitive to solve. Yet, this problem is of high relevance to many different application fields including the propagation of microwaves used for mobile communications in urban areas, the prediction of the efficiency of acoustic barriers to reduce noise levels below the thresholds for damage or annoyance, and the design of the coastal environment to handle oceanic waves, amongst the others. The best known illustrative example of the phenomenon of diffraction by a semi-infinite obstacle comes from the field of optics, being represented by coherent light passing the edge of an opaque screen [1-4]. The description of this problem involves special functions known as Fresnel's integrals [5]. In this context, and seizing on the celebration of the bicentenary of Augustin-Jean Fresnel's work [6, 7], which won the Grand Prix awarded by the French Academy of Sciences in 1819, we thought to test with a group of physics students the possible benefits of using a machine learning-based approach. Machine learning, referring to the use of statistical techniques and numerical algorithms to carry out tasks without explicit programmed and procedural instructions, is widely recognised as one of the most important technologies being developed today [8]. Machine-learning tools have been successfully used in natural language processing, translation, visual perception or anomaly detection, and applications are found in virtually every sector: e-commerce, news, media and entertainment, finance and credit, self-driving cars, and many more.

In this paper, we will first overview the physical problem that was addressed within a three-day supervised research project for the final year of the Batchelor's degree and describe the adopted approach to its solution relying on the use of a neural network. Then we will assess the network's ability to predict the observable diffraction pattern, as well as to solve the inverse problem of extracting the distance to the diffracting object from an observed diffraction pattern.

## 2. Problem under study and methods

### 2.1 Fresnel diffraction at a straight edge

Figure 1a shows the geometry of the very classical diffraction problem that was studied, i.e., the imaging of the straight edge bounding an opaque semi-infinite screen when illuminated at normal incidence by a monochromatic plane wave of wavelength $\lambda$. The diffraction pattern is observed at a distance $z$ from the screen, and $(x,y)$ and $(x',y')$ represent the transverse coordinates in the plane of the screen and the observation plane, respectively. The semi-infinite screen thus corresponds to zero optical transmission for $x<0$. This system was known to the students who had the opportunity to familiarise with it in both lecture and tutorial classes. An example of the intensity distribution in the straight-edge diffraction pattern is shown in Fig. 1(b). Ray optics would tell us that we would see a uniformly bright region separated by a sharp line from a fully dark shadow. The real situation is that light bends even into the geometric shadow, and the intensity is not constant but oscillates.

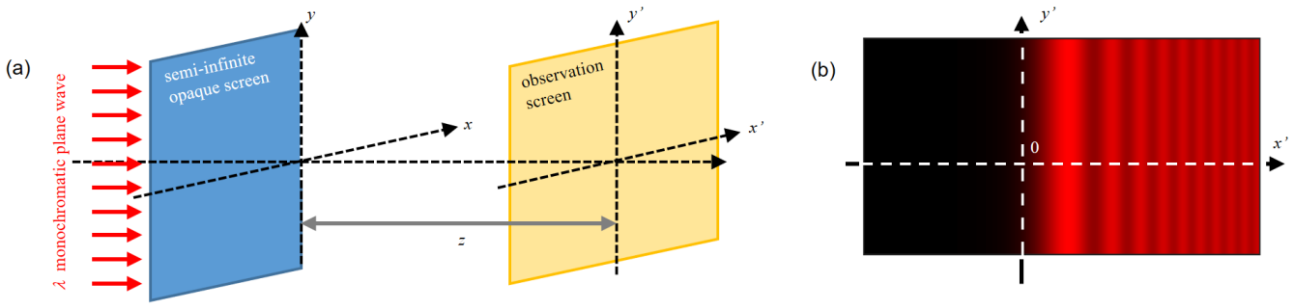

**Figure 1 – (a)** Geometry of diffraction by a straight edge. **(b)** Intensity variation near the geometric image of a straight edge (represented by a dotted white line) as observed for illumination at 632.8 nm.

The description of this diffraction problem, where the conditions used in the far-field or Fraunhofer region do not apply, requires the use of Fresnel diffraction formalism. Following Fresnel theory and considering that diffraction occurs only along the direction of the screen edge, the optical field $a(x',z)$ obtained on the observation plane is given by [1-4]:

$$a(x',z) \propto \int_0^\infty e^{i\frac{\pi}{\lambda}\frac{(x'-x)^2}{z}} dx \qquad (1)$$

where some multiplicative factors not affecting the shape of the resulting intensity distribution have been omitted for simplicity. By introducing the reduced variable

$$u = \sqrt{\frac{2}{\lambda z}}(x'-x), \qquad (2)$$

this result can be expressed as

$$a(x',z) \propto \int_{\sqrt{\frac{2}{\lambda z}}x'}^{-\infty} e^{i\frac{\pi}{2}u^2} du \propto C_f(-\infty) - C_f\left(\sqrt{\frac{2}{\lambda z}}x'\right) + i\left(S_f(-\infty) - S_f\left(\sqrt{\frac{2}{\lambda z}}x'\right)\right) \quad (3)$$

where $C_f(x)$ and $S_f(x)$ are the Fresnel integrals given by [5]:

$$C_f(x) = \int_0^x \cos\left(\frac{\pi \omega^2}{2}\right) d\omega$$
$$S_f(x) = \int_0^x \sin\left(\frac{\pi \omega^2}{2}\right) d\omega \quad (4)$$

These functions are anti-symmetric (odd) and $C_f(\infty) = S_f(\infty) = 1/2$. While Fresnel integrals are available as pre-built functions in various scientific software, it was a good exercise for the students to implement their own version using, for example, the trapezoidal rule for numerical integration. A graphical representation of the Fresnel integrals is provided in Fig. 2. Therefore, the intensity pattern that is observed on the screen is given by

$$I(x',z) \propto a\,a^* \propto \left[C_f\left(\sqrt{\frac{2}{\lambda z}}x'\right) + \frac{1}{2}\right]^2 + \left[S_f\left(\sqrt{\frac{2}{\lambda z}}x'\right) + \frac{1}{2}\right]^2 \quad (5)$$

We note that this pattern can be qualitatively predicted using the Euler spiral (also commonly referred to as clothoid or Cornu spiral) [3, 9].

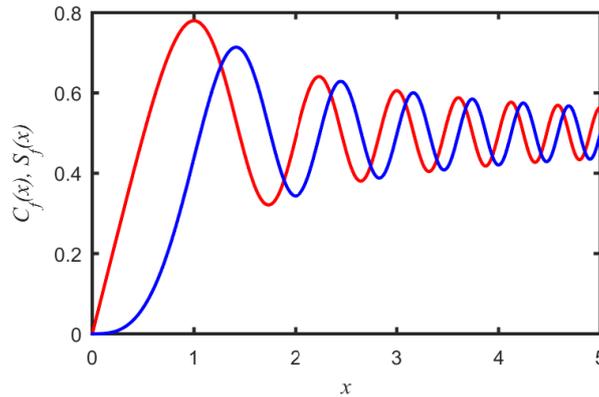

**Figure 2** – Fresnel integrals $C_f$ and $S_f$ (red and blue curves, respectively).

*2.2 Machine-learning tools*
Machine learning was a completely new subject for our students as their Bachelor's degree programme only included the basics of programming (in the Matlab language using GNU Octave). To introduce the students to the rudimentary notions of machine learning and neural networks, we made them watch the first chapter of the Massive Open Online Course on "Deep Learning" by Nicolas Thome [10], available on the France Université Numérique (FUN) online platform. In terms of software implementation, in 2018-19 we opted for the Matlab environment and its neural network toolbox [11], which has an intuitive and convenient graphical interface. In 2020-2021, we also tested the use of the Octave's package 'nnet' [12] on laptops with modest performances. Evidently, a solution in Python was fully feasible especially considering that Python is the most popular language for machine learning thanks to many available libraries and platforms.

The elementary unit in an artificial neural network is an artificial neuron, whose principle is illustrated in Fig. 3(a). The artificial neuron receives several inputs, $x_1, x_2, ..., x_n$, and sums them to produce an output (or activation) $y$. Usually each input is separately weighted by a value $w_i$, $i=1, 2, …, n$, and the sum along with a possible bias $b$ is passed through a nonlinear function $f$, known as an activation or transfer function and typically having a sigmoid shape. Hence, the output from the neuron is given by

$$y = f\left(b + \sum_{k=1}^{n} w_k\, x_k\right) \qquad (6)$$

Using the notation $\mathbf{x} = (x_1, x_2, ..., x_n)$ and $\mathbf{w} = (w_1, w_2, ..., w_n)$, Eq. (6) can be rewritten in a compact form as $y = f(\mathbf{w}\,\mathbf{x}+b)$. The learning process consists in adjusting the weights $\mathbf{w}$ and bias $b$ so to match a certain set of input parameters $\mathbf{x}$ with the target output $y$.

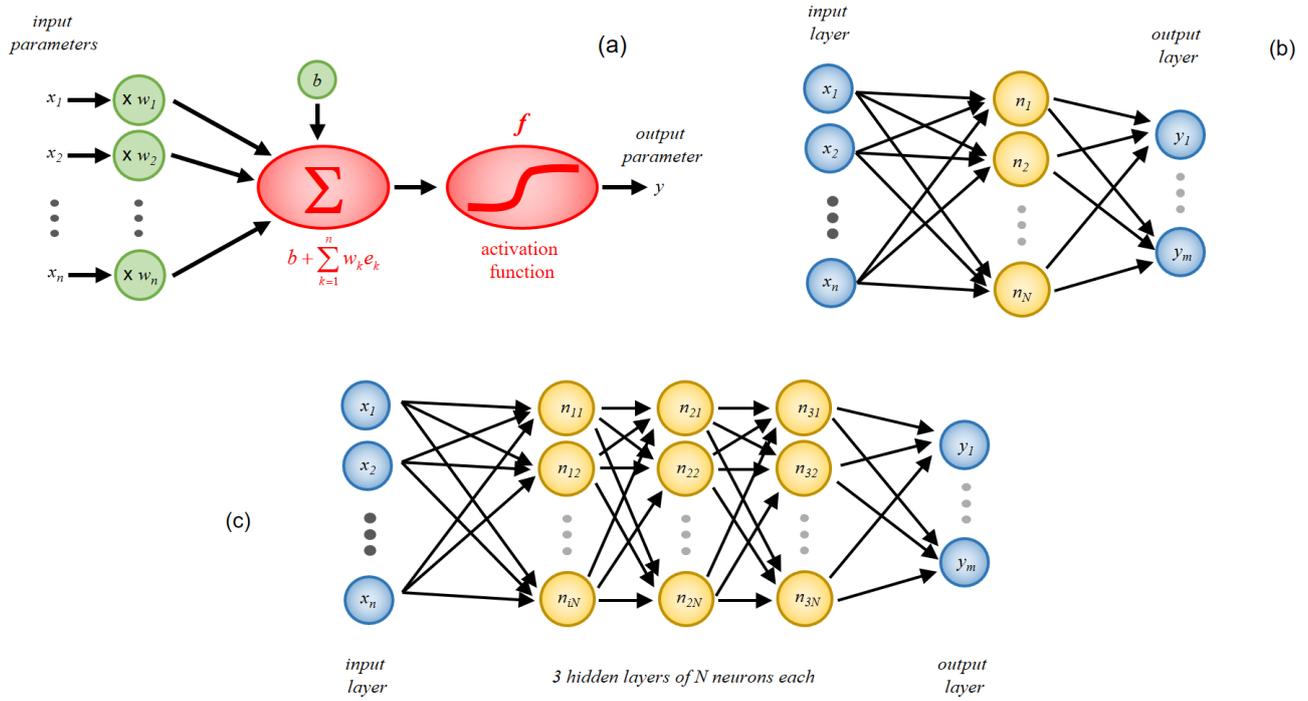

**Figure 3** – **(a)** Principle of an artificial neuron. **(b)** Feed-forward artificial neural network with a single hidden layer of $N$ neurons and **(c)** three fully connected hidden layers.

An artificial neural network is based on a collection of connected artificial neurons or nodes [10] and can produce multiple outputs. Typically, neurons are aggregated into layers, where a single intermediate or 'hidden' layer of $N$ neurons linking $n$ inputs $\boldsymbol{x}$ and $m$ outputs $\boldsymbol{y}$ constitutes the simplest network's configuration (Fig. 3b). Richer and more efficient structures can be built by spreading the neurons over multiple layers (Fig. 3c). In this case, the output $\mathbf{a}^i$ of a given hidden layer (of size $N_i$) is related to the output $\mathbf{a}^{i-1}$ of the preceding layer (of size $N_{i-1}$) by

$$\mathbf{a}^i = f\left(\mathbf{w}^i\,\mathbf{a}^{i-1} + \mathbf{b}^i\right) \qquad (7)$$

where $\mathbf{w}^i$ and $\mathbf{b}^i$ are arrays of sizes $N_{i-1} \times N_i$ and $N_i$, respectively. Powerful algorithms are then required to adjust the increased number of weighted connections so to match inputs and outputs.

## 3. Prediction of the diffraction pattern by a neural network

*3.1 Learning principles*
First, the students evaluated the ability of a neural network to predict the diffraction pattern produced by an opaque semi-infinite screen. Figure 4 summarises the various steps involved in the supervised learning process. The first step consists in creating a sufficiently large set of data to train the network and validate its predictions. Therefore, an ensemble of intensity profiles $I(x')$ corresponding to different combinations of wavelengths and observation distances were generated from Eq. (5) and regularly sampled along the spatial dimension in $m=150$ values. The input parameters to the neural network were then wavelength and distance values, $\mathbf{x} = (\lambda, z)$, and the output from the network was the predicted diffraction pattern discretised in $m$ points, $\mathbf{y} = (I_1, I_2, ..., I_m)$. The database was made of $p$ elements $(\mathbf{x}^i, \mathbf{y}^i)$. For improved network stability and modelling performance, prior to training a neural network the input and output variables must be normalised, typically in the range [-1,1] as appropriate to the scale of the activation functions. This normalisation can be transparent to the user (in Matlab for example) or explicitly programmed (in Octave). The learning methodology adopted to train a neural network is backpropagation [13]. In this method, the processed output from the network $\mathbf{y}^i_{net}$ is compared to the target output $\mathbf{y}^i$ and the computed difference or error $Err$ is propagated backwards from the output to the preceding layers and used to adjust the weight and bias values of the connected neurons. Successive adjustments will cause the neural network to produce output which is increasingly similar to the target output. The model error can be measured by the mean squared error defined as

$$Err = \frac{1}{p\,m} \sum_{i=1}^{q} \sum_{j=1}^{m} \left( y^i_{net,j} - y^i_j \right)^2 . \qquad (8)$$

where $q$ is the number of data involved in the process. After the training stage, a neural-network prediction model is usually given a dataset of unknown data against which the model is tested (called the validation dataset or testing set), where the goal of this cross-validation is to test the model's ability to predict new data that was not used in estimating it, in order to flag problems like overfitting and to give an insight on how the neural network will generalise to an independent dataset. Then the network is ready for use. Typically, we used 80% of the database for training and the remaining 20% for testing.

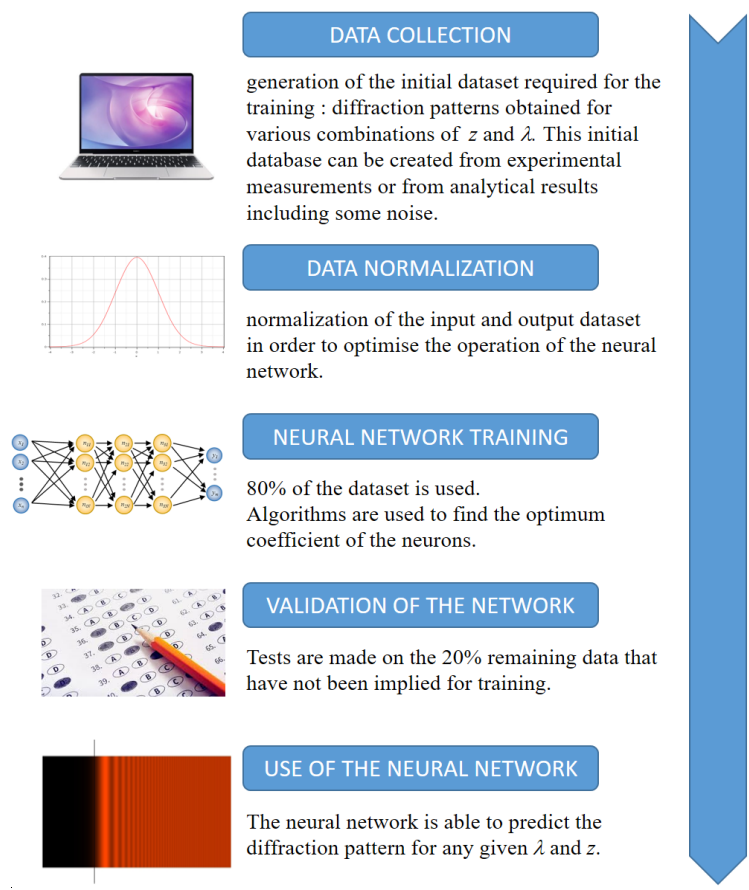

**Figure 4** – Stages involved in the supervised learning of a neural network.

Within the implementation of the learning procedure, the students had the opportunity to compare several network training functions and observe the differences in terms of prediction accuracy and speed. As discussed in the following section, they also assessed the impact of the network topology and size of the database on the network performance. After various tests, the Bayesian backpropagation and regularisation method appeared to be a robust training algorithm solution. In Octave, the available algorithm, the Levenberg-Marquardt method, also gave the most suitable results.

*3.2 Learning from ideal data*

In the first instance, the neural network dealt with ideal data, i.e., noiseless diffraction patterns as given by Eq. (5). The initial dataset consisted of the intensity profiles obtained for different wavelengths covering the visible range from 380 to 780 nm and observation distances ranging from 0.5 to 1 m. We used either a regular grid of points in the space of input parameters ($\lambda$, $z$) or randomly chosen combinations of $\lambda$ and $z$. The second approach was found to lead to better predictions by the network and, thus, was retained for the learning process.

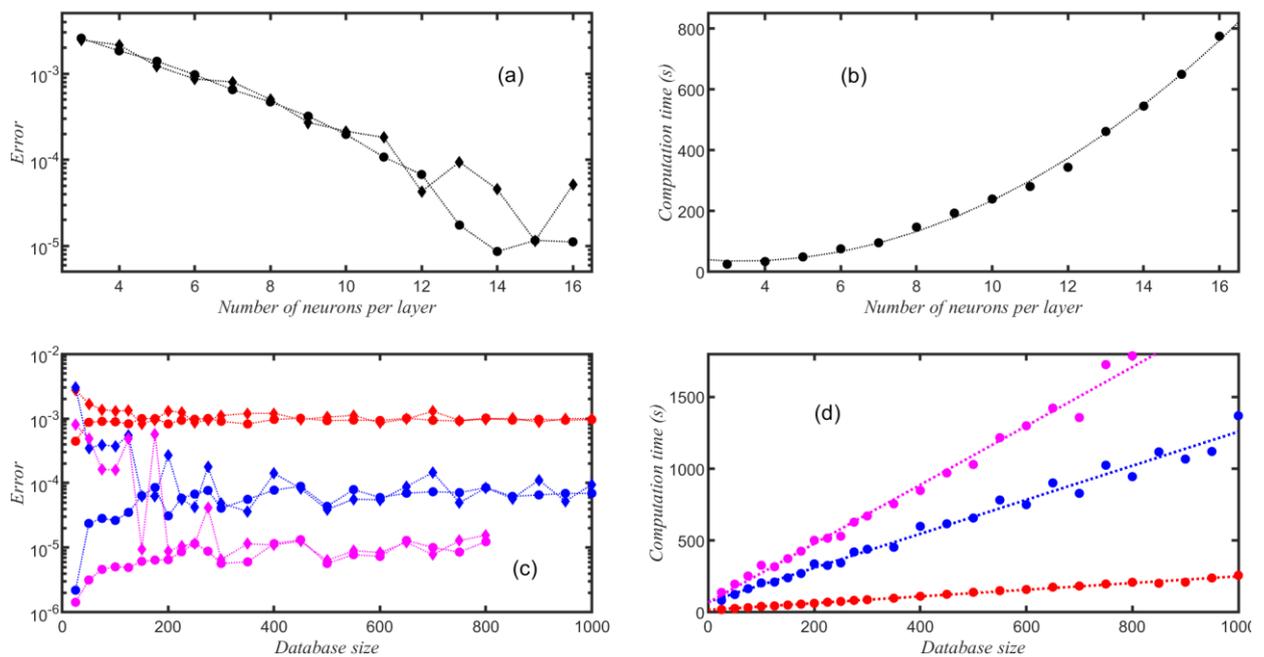

**Figure 5 –(a)** Rms errors on the training and testing data (circles and diamonds, respectively) and **(b)** training time as a function of the number of neurons per layer recorded after 100 epochs for a three-layered neural network using a database of 250 examples. The black curve in (b) is a quadratic fit. **(c)** Rms errors on the training and testing data (circles and diamonds, respectively) and **(d)** training time as a function of the database size recorded after 100 epochs for a three-layered neural network with $N = 6$, 12 and 15 neurons per layer (red, blue and purple, respectively). The dashed curves in (d) are linear fits.

We first investigated the impact of the size (number of neurons $N$ per layer) of the neural network model on its accuracy and speed when the model was trained on a set of 200 data (out of a dataset of 250 examples). Figures 5a and 5b show the root-mean-square (rms) prediction error and duration of the training phase as a function of the number of neurons per layer, recorded after 100 complete passes through the training dataset or epochs, for a network with three hidden layers. Several observations can be made from these results. First, the higher the number of neurons per layer, the better the network reproduces the data used for training, leading to a monotonic decrease in the rms error (Fig. 5a). However, this trend is not systematically observed on the unknown data used for validation, where the gain in prediction accuracy varies in a rather random fashion when the number of neurons per layer is higher than 12, thereby indicating possible overlearning of the problem and consequent overfitting of the training dataset. Furthermore, as evidenced by Fig. 5b, the learning time grows significantly with the network size, where this increase follows a close-to-quadratic law.

The size of the dataset is also an important factor influencing the network performance, as highlighted by the results shown in panels c and d of Fig. 5, which were obtained after 100 epochs for a three-layered network with 6, 12 or 15 neurons per layer. Once more, we see that the model error may vary by more than two orders of magnitude according to the network size (Fig. 5c). We also notice some overlearning instances for small datasets, where the accuracy measured against the training set is good, but the model does not generalise well to the validation set. The training time has a quasi-linear dependence on the dataset size (Fig. 5d).

In Fig. 6a we compare the diffracted intensity profiles obtained from neural networks of different sizes (6, 12 or 15 neurons per layer) for various combinations of wavelengths and observation distances. The networks were trained on 200 data. When the pair ($\lambda$, $z$) falls within the range of the data used for learning, the network predictions are in very good agreement with the analytical solution especially for the larger networks, while small discrepancies appear in the outer intensity oscillations for the smallest network (Fig. 6a1). As can be seen in the map of Fig. 6b, the prediction error remains rather small within the domain of the initial dataset, thus demonstrating the universal interpolation capability of the network model. The challenge is for the neural network to continue to make good predictions, or at least not to fail catastrophically, when the learning data covers only a limited range of the data domain. As shown in Fig. 6a2, for $\lambda = 340$ nm and $z = 1.2$ m (where this parameter pair falls outside the range of the initial data), the predictions made by the neural networks with 12 and 15 neurons per layer remain fully consistent with the analytical solution, thus suggesting that the network model can also succeed at extrapolation. However, when the observation distance is 35 cm (Fig. 6a3), none of the neural networks used can predict a fitting diffraction pattern, which reveals that the extrapolation ability of the algorithm is more precarious. It is noteworthy that in terms of the product $\lambda z$ appearing in the definition of the reduced variable $u$ (Eq. (2)), the case illustrated in panel a2 falls within the support of the training distribution whilst the case of panel a3 does not.

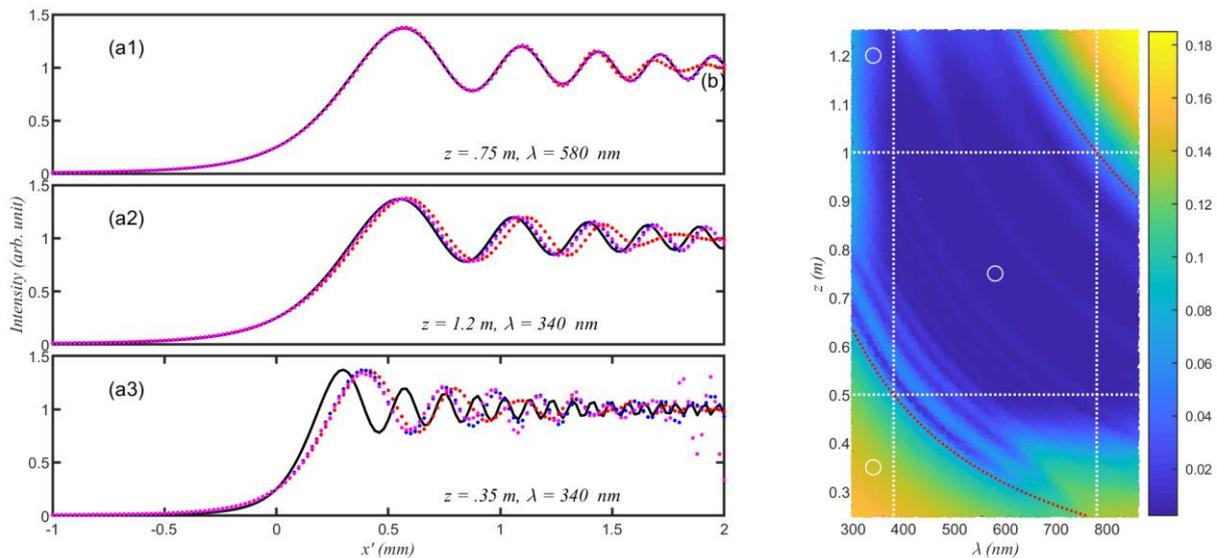

**Figure 6 –(a)** Intensities of the diffraction patterns obtained for various ($\lambda$, $z$) sets. The predictions from neural networks with 6, 12 and 15 neurons per layer (red, blue and purple curves, respectively) are compared with the analytical results from Eq. (5) (black curves). **(b)** Map of the rms prediction error of the neural network model with 12 neurons per layer in the input parameter plane ($\lambda$, $z$). The white dashed lines delimit the domain of the initial dataset. The white circles indicate the ($\lambda$, $z$) pairs used in panels a. The red dashed curves are isolines of the product $\lambda z$ calculated at the extreme values of $\lambda$ and $z$ supporting the initial dataset.

*4.3 Learning from noisy data*

At the second stage, we tested the ability of the neural network to handle noisy data. To this end, we applied to the analytical intensity profiles given by Eq. (5) additive white Gaussian noise with 10% standard deviation of maximum intensity and multiplicative noise with 10% standard deviation, which allowed us to simulate the intensity traces recorded by a 150-pixel charge-coupled device. 14 or 50 such intensity profiles, associated with observation distances in the range 0.5 to 1 m and the same wavelength of 632.8 nm (characteristic of a helium-neon laser), were used as the input data. As the dimension of the input space had decreased compared to the previous problem, we reduced the number of neurons per layer to 8 in the network model. As it is seen in Fig. 7a, despite the imperfect quality of the training dataset, the network can predict the diffraction patterns correctly, where the prediction accuracy is higher for the model trained on the larger dataset (denoted by NN2) as expected. Furthermore, Fig. 7b which shows the evolution of the diffraction pattern with the observation distance as reconstructed from the noisy data, the network predictions and Eq. (5), evidences that the neural network algorithm is even able to denoise the experimental recordings to some extent, making them more similar to the theoretical patterns. While the use of the network trained on the smaller dataset (NN1) leads to some defects in the intensity profiles at high values of $x'$, NN2 can accurately reproduce most of the theoretical features.

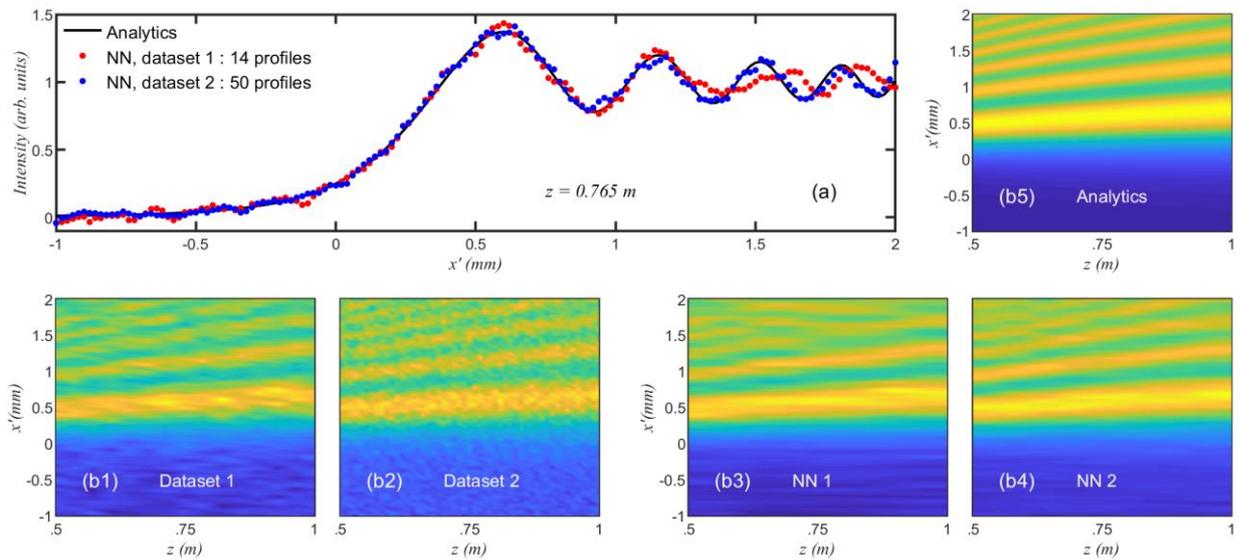

**Figure 7** – **(a)** Intensity of the diffraction pattern obtained for $\lambda$ = 632.8nm and $z$ = 0.765m. The predictions from the neural network fed noisy datasets of 14 and 50 examples (red and blue circles, respectively) are compared with the analytical results from Eq. (5) (black curve). **(b)** Evolution of the diffraction pattern as a function of observation distance reconstructed from noisy datasets of sizes 14 and 50 (panels 1 and 2), the predictions from the neural network fed 14- and 50-sized datasets (panels 3 and 4), and the theory (Eq. (5); panel 5). A false-colour scale is used to better highlight the differences among the plots.

## 4. Inverse problem: Retrieval of the system properties from a diffraction pattern

After validating the ability of a neural network to predict the diffraction pattern knowing the wavelength and observation distance, we turned our attention to the inverse problem, i.e., retrieving from an observed diffraction pattern the distance and wavelength at which the pattern formed. Such a problem required the use of a new network model where input and output had been swapped. Yet, from their very first manipulations, the students noticed that there were problems with the learning process such that the model was making erroneous predictions even when it was trained on noiseless data. It was only by looking back at the theory that the students were able to identify the origin of this failure: as shown in Eq. (2), the wavelength and observation distance play a similar role in the normalised variable $u$, where for a given diffraction pattern, there is an infinite number of combinations of $\lambda$ and $z$ values yielding the same wavelength-distance product. Consequently, and due to the underlying physics, if both parameters are unknown, it is not possible to determine them unambiguously, no matter how good the deployed neural network and quality of the dataset are.

Given this constraint, we fixed the wavelength (at $\lambda = 632.8$ nm) and tasked the trained network with the retrieval of the observation distance. The neural network used for this task included 6 neurons arranged in a single layer and, in the first instance, the training dataset consisted of 20 ideal (noiseless) intensity profiles as observed at distances between 0.5 and 1 m. We can see in Fig. 8(a) that the estimation error on $z$ (defined as the difference between the retrieved parameter value and the target value obtained from the theoretical data) is very small (well below 1mm) across the entire $z$-range covering the training data, whilst it is more pronounced at the extremities of this interval. This validates the recognition of noiseless patterns by the network model. Then the students studied the case of intensity traces affected by additive white Gaussian noise. The noise's standard deviation was 5% of maximum intensity and smoothing of the resulting intensity profiles $g$ was applied. The estimation error on $z$ was much higher in this case. The error distribution for 4000 test realisations is shown in Fig. 8b1 as it is close to a normal distribution of standard deviation 4.7 cm. Whilst the mean error is 1.2 mm and the average absolute error is only 3.9 cm, a few realisations show deviations exceeding 15 cm. Changing the network configuration seemed not to be a very effective solution to improve the results. More advanced machine-learning models should therefore be implemented when dealing with the retrieval of the system properties from experimental observations, which was well beyond the scope of our short project.

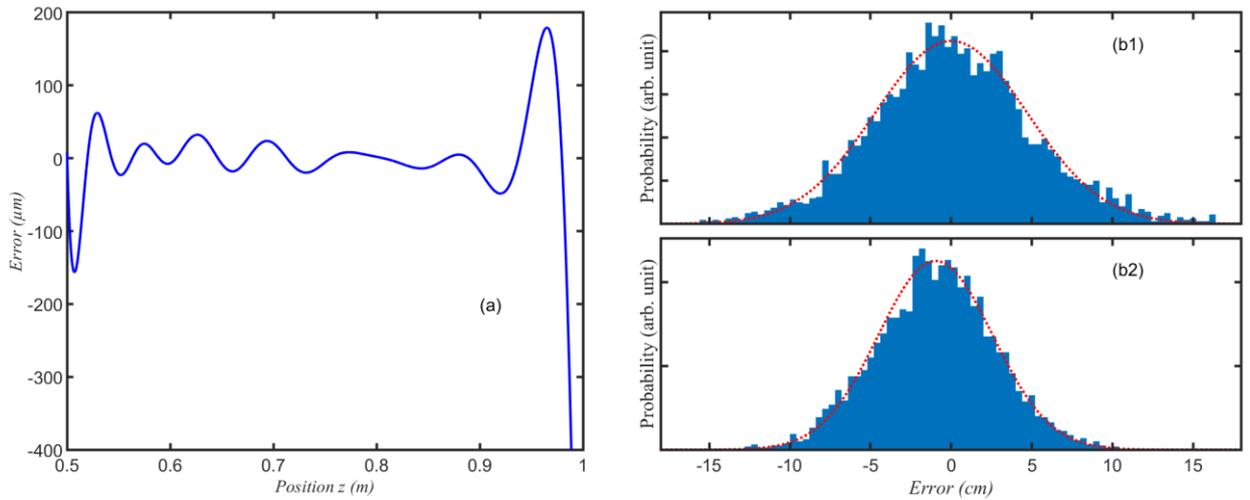

**Figure 8 – (a)** Estimation error on the observation distance $z$ when the neural network is interrogated with noiseless diffraction patterns. **(b)** Distributions of the estimation error on $z$ obtained by the analysis of noisy recordings (4000 realisations). The results obtained by use of a neural network and of Eq. (10) are plotted in panels 1 and 2, respectively. The histograms are compared to normal probability density functions (red dotted curves).

Quite interestingly, the students compared the results from the inverse-problem neural network with those obtained by use of a different and classical method which, contrarily to a neural network, requires only the position $x'_{max}$ of the first maximum of the diffracted intensity profile as the input parameter. Indeed, from Fresnel theory of diffraction and Eq. (5), it can be shown that $x'_{max}$ is given by

$$x'_{max} = \sqrt{\frac{3}{4} \lambda z} . \qquad (9)$$

hence, $z$ can be readily obtained as

$$z = \frac{4}{3} \frac{x'^{2}_{max}}{\lambda} . \qquad (10)$$

When dealing with ideal diffraction patterns, the estimation accuracy on $z$ attained with this approach was around ±1.5 cm, hence significantly lower than that achieved by the neural network. This is because the computation of $x'_{max}$ was impaired by a discretisation error resulting from the limited number of sampling points considered in the intensity profile. Note that this discretisation error could be minimised by implementing some form of interpolation in the vicinity of the maximum of the intensity profile. Conversely, when dealing with noisy diffraction patterns, the 'old-fashioned' method outperformed the neural network model, as evidenced by the distribution of errors shown in Fig. 8b2.

## 5. Conclusion

We have reported on a research project carried out over a few days with physics undergraduate students with no prior knowledge of machine learning, which allowed the students to get acquainted with the basics of machine learning by using a known optical problem as a testbed. Without going into details about the algorithms involved and without optimising the techniques chosen (the use of a convolutional neural network would be a plus), the students could appreciate the usefulness of neural networks for the prediction and recognition of a diffraction pattern. They could also understand that sometimes performance limitations are not related to the machine learning algorithm used but rather to the underlying physics, hence they require understanding of the phenomenon being modelled. In other words, machine learning does not necessarily compete with traditional approaches. Further, they recognised the crucial role of the input data in predictive modelling.

The students completed the project by reading some research papers published by the supervisory team on the use of neural networks for the modelling of advanced optical systems [14-17] where no analytical formulation is possible. Indeed, the application of machine learning techniques is transforming the scientific landscape, yielding new insights into many areas of fundamental and applied science [18]. Photonics is no exception, and recent years have seen the rapid growth and development of the field of smart photonics, where machine-learning algorithms are being matched to optical systems to add new functionalities and to enhance performance [19-22].

The students presented the project outputs in the form of an individual eight-page report. Piloted with three students in 2018-2019, this project was proposed again to two students in 2020-2021. The main outcome from the project was a demystification of machine learning and a better understanding of the importance of big data in data-driven decision-making algorithms. The students also realised that, starting from suitable libraries, the implementation of a multilayer neural network requires only a few lines of coding. One of the main issues raised by the students was about the physical mechanism that makes a neural network work. They finally understood that supervised learning algorithms such that used in this project infer a function that maps an input to an output from example input-output pairs, and do not intrinsically involve any physics. A rough analogy with polynomial interpolation helped them grasp the very different natures of a simple neural network and a well-established physical theory.

It is envisaged that this three-day project, where students explore and test their ideas, will partly evolve into a four-hour practical session for the whole final-year student cohort. The concepts of convolutional neural networks and deep learning will also be discussed, together with examples where physical modelling is inferred from the use of machine learning strategies [23].

**Acknowledgements**
We sincerely thank Leo Théodon, Nicolas Thevenot and Guillaume Jaskula, students of the Bachelor's degree programme in Physics at University of Burgundy, for their commitment to this project and the high quality of the work produced. We also acknowledge support from the French National Research Agency (I-SITE BFC ANR-17-EURE-0002 and OPTIMAL ANR-20-CE30-0004 projects).